\documentstyle[11pt]{article}
 
\def\pa{\partial}
\def\k{\kappa}

 \def\D{\Delta}

\def\k{\kappa}
 
\def\m{\mu}

\def\mn{{\mu \nu}}

\def\be{\begin{equation}}
\def\ee{\end{equation}}
     
\begin{document}

\begin{flushright}
BRX TH--459
\end{flushright}

\begin{center}
{\Large\bf Two Outcomes for Two Old
(Super-)Problems}\footnote{Dedicated to two founders of
D=4 SUSY on its (nearly) 30$^{\rm th}$ birthday.}

\vspace{.4cm}

{\sc S. Deser}

\vspace{.3cm}

{\em Department of Physics} \\
{\em Brandeis University, Waltham, MA 02454, USA} 

\vspace{.2cm}
\end{center}

\begin{quotation}
{\bf Abstract}:
I briefly review the outcomes of two very different old 
questions -- where global SUSY improves on ordinary QFT -- 
when they are in turn posed in the local, SUGRA, context.  The
first concerns the unexpectedly powerful role of the 
``Dirac square root" graded
algebra relation $E= Q^2$, originally found
in D=4 SUSY by Gol'fand and Likhtman, in proving
positive energy theorems both in SUGRA and in ordinary
gravity theories, where $E$ and $Q$ have very different --
gauge generator -- definitions.  The second seeks SUGRA
counterparts of the boson/fermion loop ultraviolet cancellations
in SUSY models. Here the result is negative: local
supersymmetry cannot overcome the infinities associated
with the dimensional gravitational $\k^2$ even,
as recently shown, in maximal D=11 SUGRA.
\end{quotation}

\noindent{\bf 1. Introduction}

We are here commemorating the birth of D=4 supersymmetry,
(SUSY), as well as the memory of the late Yuri
Abramovich Gol'fand, who with 
Evgeny Pinkhusovich Likhtman, discovered it \cite{001}
in one of those Soviet-era works often overlooked in the 
West -- and in this case, also in the East!

So universal has the SUSY gospel become in so short a
time that few of us still envisage particle theory
without (broken) SUSY or gravity without supergravity
(SUGRA), even in the absence of any experimental
evidence.  This is a strong testimonial to the
naturalness of SUSY, since it differs from ordinary 
QFT in a ``discrete" way.  By this I mean that a
generalization, such as Einstein's of Newtonian gravity,
is usually continuously related to it by a natural group
contraction, there $c \rightarrow \infty$.  Here there
is no such parameter (except the far too strong
$\hbar \rightarrow 0$) to rely on.  Nor do we have a unique
natural SUSY-breaking mechanism.  While awaiting explicit
verification of uniquely SUSY predictions, its theoretical
persuasiveness rests on tangible formal successes.  In this
essay, I will concentrate on two problems where SUSY
undeniably improved on ``ordinary" QFT, but take them one
step further, translated to the local SUSY, namely SUGRA,
context.  I chose the two for balance: 
one succeeded beyond expectation,
the other failed according to expectation.

The first problem stems from perhaps the 
deepest aspect of the graded
Poincar\'{e} group, the supercharge anticommutator
given  in \cite{001}, whose relevant part for us is that
the square of two supercharges is
the energy.  I will recall the adaptation of this
relation to the proof of
positive energy of supergravity and of Einstein
gravity, where both supercharge and energy have very different
meanings.  My other example involves heavier machinery, and
in its latest and final incarnation the result has only 
recently been obtained:  what is the impact of supersymmetry on
ultraviolet divergences in SUGRA?  Are any supergravity
models renormalizable, or more exactly, finite?  Since this 
is neither a report of new work nor a real review,
all details (and many references) are left to the 
references.

\noindent{\bf 2. Positive Energy}

If one looks at the graded algebra of SUSY, it contains
(besides the existence of a constant supercharge)
one really novel relation, that
the square of
$Q$ is the total energy.  This is the true
``improvement" on the Poincar\'e algebra, imposing the
huge (and desirable) restriction on 
supersymmetrizable models, that more generally,
their 4-momentum is future
timelike. I should caution that it does
{\it not} exclude systems having a trivial kind of negative
energy.  Consider for example, two copies of a free multiplet,
the second of whose actions has relative ghost sign; each is
separately supersymmetric and yet the total energy need not
be positive.  Indeed, the more refined arena of SUGRA
models with actions quadratic in curvature is 
essentially of the above type, and provides some 
instructive examples \cite{002} of the role of the 
Hilbert space metric in interpreting 
``$E=Q^2$".  Still,
this caveat is more of a curiosity; every 
bose/fermi pair of excitations must share
purely positive (or purely negative) energy, and that is
really the heart of SUSY's ``Dirac square root" 
character, as was already exploited
(for global SUSY) in \cite{zumino}.

In 1976, when SUGRA was found, the
problem of positive energy in Einstein gravity (GR) had
more or less exhausted those of us who had worked on it
over the previous 15 years, ever since gravitational energy
first became sufficiently understood to ask the question.
While there were no (correct) 
counter-examples, there were also only ``physicists' 
proofs" \cite{004}. It was only for
special initial value configurations that explicit
positivity could be exhibited,
and no underlying ``square of a square root" structure
of the energy was ever observed in the process. It was
agreed (among some of the experts at least) that 
it was both necessary for physical consistency 
and probably true, that not only did GR have positive 
energy  but that 
$E=0$ implied the vacuum, {\it i.e.}, only flat space 
had vanishing energy (for some history, see \cite{006}). 
Actually, just about that time the 
first positivity
proof did emerge \cite{005}; while rigorous, it was
not very intuitive. The obstacle to obtaining a direct
physical proof of positivity of GR in the sense that 
the energy of
say (even nonabelian) vector gauge
theory is manifestly positive, $E = \frac{1}{2} \int  d^3r 
(\mbox{\boldmath $E$}^2 + 
\mbox{\boldmath $B$}^2)$, is that not only is
GR a gauge theory with energy as the ``charge", but an 
infinitely self-interacting one. While its (linearized)
free massless spin 2 excitations
consist of a positive harmonic-oscillator frequency sum
once the gauge constraints are imposed,
the higher terms are not easily bounded: 
the full Einstein theory's action is an infinite series
$I_E \sim \int d^4x \sum^\infty_{n=1} h^n (\pa h)^2$,
where $h_\mn = g_\mn - \eta_\mn$ is the deviation from
flatness of the covariant metric, on top of which 
the ten $h_\mn$
consist of four pairs of unphysical gauge and 
constraint variables (akin to the longitudinal vector
potential and electric 
field in electrodynamics), the latter are themselves spatially
non-local functionals of the two remaining physical components
(as well as of the matter stress-tensor).

Aside from the above technical difficulties
of the pure GR problem,
interest now moved
to the more complicated -- because including a ``matter"
source and being intrinsically quantized -- 
SUGRA, seemingly an even more difficult arena.
Yet it had that magic $E=Q^2$ formula;
however, when we turned it 
to advantage  \cite{007}, we were met with scepticism
because of the well-known special status as 
surface integrals
of $E$ already in GR, and $Q$ for the spin 3/2 field.  
Besides, how could an ancient unsolved problem
in the simpler classical setting be trivially
solved in this much more complicated one? This is 
not the place to provide the details of our results,
but only to indicate the beauty of the theory and 
the robustness of these ``charges" upon transition to
the SUGRA context.  Recall that in GR -- with or without
sources such as the spin 3/2 field $\psi_\m$ -- the total
energy $E$ (defined for asymptotically flat solutions)
being a gauge charge, could be expressed as a surface
integral of a purely
gravitational quantity at spatial infinity.  
To oversimplify enormously what took
a long time to understand, the equivalent of the 
Gauss constraint on the Coulomb
potential,  with charge density as source
$(\nabla^2 \varphi = \rho )$ is here 
the $G^0_0 = T^0_0$ Einstein constraint, which
can be rewritten to resemble
linear Gauss form for a certain component
$h^T$ of the metric, $\nabla^2 h^T = t^0_0$.  Here,
however the energy density $t^0_0$ 
of gravity plus matter depends in a complicated way
on $h^T$: energy self-interacts!
Still, the value $E$ can be read off by elementary
Poisson equation considerations as the coefficient of
$1/r$ in $h^T$ at spatial infinity (whatever the complications).
But whether its sign
is always positive
remains hidden in the full interior ``counting",
$E = \int d^3x t^0_0 (h)$, of the local contributions,
just as the total charge's functional form, 
in contrast with its surface value, as the coefficient
of $1/r$ in $\varphi$, is only obtainable
by ``counting" all the interior electrons.  Fortunately,
if it can be shown generically that $E$ is
itself always the square of a Hermitian
operator, then this suffices 
to prove its positivity by functional form
without needing recourse
to the volume integral.  This is where the supercharge
$Q$ of SUGRA comes to the rescue.  Everything that has
just been said in our child's description
of gravitational energy holds for the spin 3/2 field's
supercharge.  It is also simultaneously both a
surface integral as well as
one over the interior volume.  What we were then able
to show was that indeed the Gol'fand--Likhtman relation
$E = Q^2 \geq 0$
continues to hold as an operator statement
in SUGRA, where $E$  is now the total
(non-vanishing!) Hamiltonian operator
of SUGRA, including both graviton and gravitino
contributions. [This is an essential point, since the
$(E,Q)$ seemingly vanish by the respective
constraint equations, and a delicate analysis
is required.]  Furthermore,
because the relevant Hilbert space (after gauge fixing)
has positive metric, $E$ can vanish if and
only if $Q$ vanishes on all states; but
$Q|>=0$ in turn
implies that
there are no excitations: spacetime is then flat and there
are no gravitinos.  This result even remains valid
in presence of supermatter 
(and hence necessarily of positive energy)
sources, since $(E,Q)$
remain the total generators, and obey the
same algebra independent of any sources.
Those details are simply hidden in the constraints.  Likewise,
for $N > 1$ SUGRAs, the $E$-$Q$ relation still holds as a
sum over the various charges, $E=\sum^N_{i=1}
Q^2_i \geq 0$.

The positive energy proof
was perhaps the first formal benefit that 
SUGRA brought to the 
gravity world, but at first it seemed both too much and
too little.  The ``too much" was that it was a
{\it quantum} result; what was its implication to
the ``smaller" world of classical and non-SUSY gravity, with
sources such as stars or photons?  
The ``too little" was that formal results of SUGRA could not
be trusted, since its likely nonrenormalizability 
(see Sec. 3) would render all
quantities infinite and the relations between them suspect.
Given that the magic $E=Q^2$ mantra yielded so much for so 
little effort compared to the frontal assaults, both  
unsuccessful and successful, 
in the classical theory, we cautiously claimed only the
formal SUGRA victory in \cite{007}.  
It was soon pointed out, however, that
at this same formal level our proof remained valid in the
classical and non-fermionic limit \cite{008}:  as
$\hbar \rightarrow 0$, all
internal, closed loop, contributions from either the 
graviton or from $\psi_\m$ disappear, while considering matrix
elements of the $E=Q^2$ operator relation between purely bosonic
states then reduces everything, including the result, 
to ordinary tree-level/classical GR.

The next step was to find a purely classical proof
that utilized only the ``smile" of SUGRA, {\it i.e.},
the necessary positiveness of pure GR that
underlies its supersymmetrizability.
The ``Dirac square" character
of the classical GR constraint equations that define the
four-momentum for asymptotically flat space was
exhibited, in a celebrated work \cite{009}, 
in terms of a classical spinor parameter that is the 
sole (gauge function) remnant of SUGRA; this was
refined and spelled out further in many subsequent papers
({\it e.g.}, \cite{deser}).
So the SUGRA argument really did teach us a lot about the 
physical constraints imposed 
on a classical theory, in this case GR,
for it to be part of a SUSY one, thereby also providing
yet another lesson in
the liberating effects of (Grassmann) extending number
systems.  There have been many subsequent examples of 
SUSY constraints, from the restrictions on
potentials allowed in normal SUSY models to the simple 
helicity-conservation rules dictated for
graviton-graviton scattering in GR due to its
SUGRA embeddability.  [There is undoubtedly still progress
to be made in this connection at higher dimensions, by
finding the right D$>$4 helicity generalizations and 
thence their SUSY constraints.]  Let me also
mention that the currently important world of ADS gravity,
with negative cosmological constant, also benefits from
the above energy results.  For, not only can SUGRA be extended
to encompass anti-deSitter sign cosmological terms (except
for the maximal, D=11, case \cite{011}) but so can the
notion of energy -- or at least the nearest that the
asymptotically ADS (rather than flat) boundary conditions
allow there \cite{012}.

\noindent{\bf 3. Nonrenormalizability of SUGRAs}

My second, very different, topic 
has been studied almost as long as SUGRA itself; its 
renormalizability (or more precisely, finiteness; the
dimensional nature of the gravitational constant precludes
counterterms of the same derivative order as the original action).
Although not discussed directly in \cite{001}, it was soon
realized, in the realm of ordinary SUSY theories, that
having equal numbers of fermionic and bosonic excitations
leads to all sorts of marvellous cancellations of ultraviolet
loop divergences, such as that of zero-point energy \cite{zumino}.
It was no wonder then, that one of the hopes following 
the discovery of D=4 SUGRA was that, despite 
describing gravity plus (spin 3/2) ``matter"
(which was already known to be one-loop divergent generically),
it would be more convergent by virtue of its extended
symmetries.  Indeed, there was both one- and
two-loop improvement: The 
one-loop counterterms, precisely as in pure gravity
(rather than as in gravity coupled to ``ordinary" matter),
were ``field-redefinable-away" arrays proportional to the
field equations of supergravity, {\it e.g.},
$\Delta I_1 = \int d^4x (G_{\mu\nu} - \kappa^2
T_{\mu\nu} (\psi )) X^{\mu\nu}$,
while the two-loop term (where GR first fails \cite{Sagnotti})
was absent altogether due to supersymmetry,
there being no SUSY companion to
$\kappa^2R^3_{\mu\nu\alpha\beta}$.
Alas, \cite{013}, three-loop invariants 
did exist both for the original N=1 and higher $N$ models. 
Their structure follows the lines of the simpler global 
SUSY case, where
the system's stress tensor $T_{\mu\nu}$, supercurrent
$J_\mu$ and chiral current $C_\mu$ form a 
supermultiplet from which a quadratic SUSY invariant
$\sim \int d^4x [T^2_{\mu\nu} -i \bar{J}_\mu \not\!\partial J^\mu
+ \frac{3}{2}C_\mu \Box C^\mu ]$
can be  constructed.  As we know,
there is no tensorial
stress-tensor for the gravitational field itself, the
nearest, higher derivative, analog being
the Bel--Robinson tensor 
$B_{\mu\nu\alpha\beta}$ quadratic in the curvature.
On-shell $(R_{\mu\nu} = 0)$ $B$ is totally symmetric,
(covariantly) conserved and traceless and there is again a
quadratic invariant 
$\sim \int d^4x [B^2 -i \bar{J} \not\!\partial J
+ \textstyle{\frac{3}{2}} C \Box C ] $
constructed from $B$ and corresponding
(also higher derivative) super- and chiral
currents. Surprisingly, 
it was even possible to learn something about the
actual coefficients of such counterterms:  For the non-maximal
1$\leq$N$<$8 models, where covariant superspace quantization
is possible, it was concluded \cite{014} that they
are uniformly non-vanishing.
For the special maximal N=8 case, very beautiful recent
work \cite{015} suggests that, while the three-loop coefficient may
vanish, its 5-loop coefficient probably does not. Similarly, 
for dimensions, 4$<$D$<$11, 
it is possible to construct
SUSY counterterms explicitly where a suitable superspace
formulation exists and the construction and negative
conclusions of \cite{015} apply.

This left only D=11 SUGRA \cite{016} as the remaining
candidate to which the earlier arguments did not
directly apply.  After a long period of neglect because
it fell beyond string theory's D=10 regime, it has revived
in the wake of its M-theoretical connections.  It
represents the highest-dimensional 
SUSY model allowed by
kinematic consistency (a single graviton, no
higher spins etc.).  It is unique
in many other respects, including 
being only N=1, {\it i.e.},
simultaneously ``minimal/maximal" and matter-free because
there is no lower spin SUSY system at D=11. Unfortunately, 
it also lacks any formalism that would enable one
to test the SUSY of, let alone construct,
candidate invariant counterterms.  
It is this 
construction and its consequences that I now describe;
details are in \cite{017}. The 
celebrated D=11 SUGRA action has, in its bosonic sector,
in addition to the graviton,
a 3-form potential $A_{\mu\nu\alpha}$
with associated field strength
$F_{\mu\nu\alpha\beta} \equiv \partial_{[\beta}
A_{\mu\nu\alpha ]}$.  It contains besides their
kinetic terms, a cubic metric-independent 
Chern--Simons (CS)
term in which the 11 indices of the epsilon symbol
are saturated by two $F$'s and one $A$.  
It is clear, because dimension is odd
(and $\k^2 \sim L^{+9}$ since the
Einstein action here is $I_E = \k^{-2}
\int d^{11} \sqrt{-g} \, R$), that
no 1-loop $\sim \kappa^0 \int d^{11}x$ candidate
$\Delta I_1$ exist -- one cannot make gravitational
scalars with odd numbers of derivatives.  At 2-loop,
lowest possible, order then, 
$\Delta I_2 \sim \kappa^2 \int d^{11}x \, \Delta L_{20}$
where $\Delta L_{20}$ has dimension 20.
The dimensional
regularization we use here uniformly has logarithmic
cutoff so all $\Delta L$ are $\Delta L_{20}$'s.
Now since a curvature is
dimensionally equivalent to two covariant
derivatives $D_\mu$, candidate terms are
schematically of the form $\Delta L_{20} \sim
\sum^{10}_{n=4} R^n \, D^{2(10-n)}$.  
The simplest and first interesting level is $n=4$
because clearly the lower $n$'s are either
(like $R^3$) not parts of super-invariants or are leading
order trivial (like $R^2$ which can be removed
by field redefinitions).  Thus, our
lowest possible choice is $\Delta I_2 \sim \kappa^2
\int d^{11}x [R^4 D^{12} + \ldots ]$ where the
ellipsis represents the SUSY completion, if any.
How does one construct a suitable $\D I_2$ in absence of any
guiding super-calculus?  Our procedure was the
following.  As was also recognized in \cite{015},
there is certainly one on-shell nonvanishing
lowest order SUSY invariant that starts out quartic
in $h_{\mu\nu}$: the tree-level 4-point
scattering amplitude M generated by the D=11 action
itself.  It has the enormous advantage that,
since there are no loops, and SUSY transformations are
linear at this level, the purely bosonic terms are
guaranteed to be part of the overall SUSY invariant that
is the total 4-point amplitude.  However, its
usefulness in providing a counterterm faces
two immediate obstacles:  We want a local
invariant and indeed one with twelve explicit
derivatives, whereas the amplitude has a denominator, from
virtual particle exchanges; can we extract such a local 
but still SUSY residue L from M?  The answer is yes,
because each term in the amplitude is in fact 
proportional to the product $(1/stu)$ of the Mandelstam
variables and we can insert the additional $D^{12}$,
without losing SUSY or having everything
vanish on-shell, by
further multiplication with $(stu)^2$ or 
$(s^6 + t^6 + u^6)$, after the initial $stu$ one.

After expanding the Einstein plus form field actions to
lowest relevant order in $h_\mn$, one computes the
various 4-part amplitudes. These consist of four
different types of scatterings:
a) graviton-graviton ($\sim R^4$), b) graviton-form
($\sim R^2 F^2$), c) form-form ($\sim F^4$) 
and d) form-graviton
bremsstrahlung ($\sim F^3R$). It is a straightforward, 
if index-intensive, procedure to
perform the explicit calculations.  The worst, graviton-graviton
scattering, is fortunately already known
\cite{018} and provides a very useful check. 
It is proportional to
the famous lowest string correction to the Einstein action
in D=10: $L_g = stu \: M_g \sim t_8 t_8 \: RRRR$
where $t_8$ is a constant 8-index tensor.  The 
three remaining, form-dependent, 
types of amplitudes can also be calculated. 
[The resulting total bosonic 
component of the full SUSY invariant
is interesting in its own right as the 
correction to the D=11 SUGRA action from M-theory, of
which we mostly know that it contains SUGRA
as the local limit.]  Let me summarize the 
various ``localized" on-shell 4-point
amplitudes a)--d).  
Schematically (see \cite{017} for details), with $B$ a
Bel--Robinson-like curvature quadratic and $F$ always
appearing with a gradient,
$ \Delta I^B_2 (g,F) = \kappa^2 \int d^{11}x 
[B^2 + (\partial F)^4 + B(\partial F)^2
+R (\partial F)^3]$, where we can actually write each part
as squares of (themselves quadratic) currents.
These terms
are at this stage
only accurate to lowest, fourth, order in the
combined $R$'s and $F$'s, and of course are to be 
supplemented by fermion-dependent terms that we
do not write down.  Nevertheless linearized SUSY is
guaranteed by our construction, and the existence of a full
on-shell invariant (however horrible) is also secure.

Of course, a counterterm is only dangerous if its
coefficient is non-zero.  A powerful tool for 
answering this question was provided by the amazing 
correspondence between SUGRA and Super-Yang--Mills 
models established and exploited in \cite{015}
to obtain many otherwise ``impossible"
results, such as mentioned above for
N=8 D=4 SUGRA.  While SYM is only defined for
D$\leq$10 one can argue, quite convincingly,
that the results as provided really do not depend directly
on D, and extend analytically
also to D=11.  This extension may also be possible to verify
intrinsically; thus,  already at leading
2-loop level,  the strong odds are against finiteness
of even this maximal theory.

\noindent{\bf Summary}

In this celebration of the birth of D=4 SUSY, we have discussed
some consequences
of the supersymmetry algebra and of its invariants within
the local, SUGRA, extension.  The  core of SUSY
grading is $E=Q^2 \geq 0$ and
requires candidate SUSY models to have positive energy, up
to some (noted) fine print.  This seemingly banal 
requirement on supersymmetrizable QFT
had an enormous impact on SUGRA and thence on
pure GR because the supersymmetrizability-Dirac square
root relation yielded an immediate physical understanding
of positive gravitational (as well as SUGRA)
energy.  The second effect of SUSY, to soften
ultraviolet divergences by producing cancellations in
boson/fermion loop contributions,
could not, however, overcome the dimensionality of the
Einstein constant even in ``maximal" D=11 SUGRA: There exist
invariants that can provide counterterms and their coefficients do not
vanish.  This negative result points
to the need for string-like
(but still SUSY) nonlocality to provide a finite 
description of gravity; a fitting conclusion, given the 
still more elemental, D=2, source of supersymmetry!

\noindent {\bf Acknowledgements}: This work was supported by
the National Science Foundation under grant PHY99-73935.
It is a pleasure to have collaborated respectively
with C.\ Teitelboim and  D.\ Seminara on the joint
work reported here.

\end{document}